\documentclass[paper,prd,reprint,preprintnumbers,nofootinbib,notitlepage,showpacs,showkeys]{revtex4-1}
\usepackage{latexsym,graphicx,amssymb,amsmath,mathrsfs} 

\DeclareSymbolFont{bbold}{U}{bbold}{m}{n}
\DeclareSymbolFontAlphabet{\mathbbold}{bbold}

\newcommand{\be}{\begin{equation}}      
\newcommand{\ee}{\end{equation}}      
\newcommand{\bea}{\begin{eqnarray}}      
\newcommand{\eea}{\end{eqnarray}}    
     
\newcommand{\rt}[1]{{}}

\newcommand{\Tr}{\,\textrm{Tr}\,}

\newcommand{\ife}{\,\textrm{if}\,}
\newcommand{\orr}{\,\textrm{or}\,} 
\newcommand{\els}{\,\textrm{else}\,}

\makeatletter
\renewcommand\appendix{\par
\setcounter{section}{0}%   
\setcounter{subsection}{0}% 
\gdef\thesection{\appendixname\space\@Alph\c@section}}

\long\def\unmarkedfootnote#1{{\long\def\@makefntext##1{##1}\footnotetext{#1}}}
\makeatother

\begin{document} 
%{\allowdisplaybreaks

\title{Functional dependence of axial anomaly via mesonic fluctuations \\ in the three flavor linear sigma model} 

\author{G. Fej\H{o}s}
\email{fejos@riken.jp}
\affiliation{Theoretical Research Division, Nishina Center, RIKEN, Wako 351-0198, Japan}
\preprint{RIKEN-QHP-192}

\begin{abstract}
{Temperature dependence of the $U_A(1)$ anomaly is investigated by taking into account mesonic fluctuations in the $U(3)\times U(3)$ linear sigma model. A field dependent anomaly coefficient function of the effective potential is calculated within the finite temperature functional renormalization group approach. The applied approximation scheme is a generalization of the chiral invariant expansion technique developed in [G. Fej\H{o}s, Phys. Rev. D {\bf 90}, 096011 (2014)]. We provide an analytic expression and also numerical evidence that depending on the relationship between the two quartic couplings, mesonic fluctuations can either strengthen or weaken the anomaly as a function of the temperature. The role of the six-point invariant of the $U(3)\times U(3)$ group, and therefore the stability of the chiral expansion is also discussed in detail.}
\end{abstract}

\pacs{11.30.Qc, 11.30.Rd}
\keywords{Axial anomaly, chiral symmetry breaking, functional renormalization group}  
\maketitle

\section{Introduction}

Axial anomaly is the anomalous breaking of the $U_A(1)$ subgroup of approximate $U_L(N_f)\times U_R(N_f)$ chiral symmetry of quantum chromodynamics. It can be understood theoretically through instanton solutions of the classical equations of motion describing vacuum to vacuum amplitudes with different topological winding numbers \cite{schaefer98,fukushima10}. Although the origin of the anomaly has been clarified for a long time, very little is known about its finite temperature restoration, especially around and below the critical point. It is well established that due to disappearing instanton density at high enough temperature or chemical potential the $U_A(1)$ anomaly has to vanish \cite{thooft76,gross81}, however, it is an open question how relevant it is at the chiral transition point. There are recent experimental findings that show a reduction of the $\eta'$ mass near the chiral crossover temperature \cite{vertesi11}, which might be related to the restoration of the $U_A(1)$ factor around $T_C$.

The relevance of the anomaly around the critical point is crucial from the point of view of the order of the chiral transition itself, since the effective $N_f$ flavor low energy description provided by the $U(N_f)\times U(N_f)$ linear sigma model \cite{gellmann60,paterson81} predicts a first order transition for $N_f\geq 2$, if the anomaly is restored at $T_C$ and no explicit symmetry breaking terms are present. This is based on a fixed point analysis of the renormalization group (RG) flows \cite{pisarski84}, and also on explicit calculations of the effective potential \cite{fejos14,fukushima10b}. Even though there are indications that the RG argument might not survive for $N_f=2$ \cite{pelissetto13,grahl14,nakayama14}, it seems to remain correct for $N_f\geq 3$, which includes the most important $N_f=3$ case.

Beyond the anomaly free fixed point analysis of the aforementioned $U(N_f)\times U(N_f)$ scalar model, it can be easily extended with a $U_A(1)$ breaking term, also known as the 't Hooft determinant. Its coupling is usually chosen to be proportional to the topological susceptibility ($\chi_{top}$), in order to reproduce the Witten-Veneziano relation (WVR) \cite{witten79,veneziano79}, and hence shows an exponential suppression as a function of the temperature. It is based on linking $\chi_{top}$ to the instanton density at finite temperature, which, if the tunnelling amplitude is estimated by a semiclassical approximation, shows an exponential damping. However, this is controversial since strictly speaking, the semiclassical approximation is applicable only for temperatures higher than $T_C$. There have been attempts to use a modfied WVR \cite{benic11}, but if the temperature dependence of the `t Hooft coupling is moderate below $T_C$, one is interested in investigating the role of mesonic (both thermal and quantum) fluctuations on the $U_A(1)$ factor, using a temperature independent coupling. 

The goal of this paper is to develop a new method that is able to give an account of the effect of the mesonic fluctuations on the anomaly. This will be achieved by generalizing the chiral invariant expansion \cite{fejos14} of the effective potential in the functional renormalization group (FRG) formalism. 

FRG and related approximations have been proven to be valuable tools in understanding the phase structure of scalar theories \cite{wetterich93,berges02,pawlowski07,blaizot06}. It has been shown in several studies that, even the leading order of the derivative expansion applied at finite temperature gives decent results \cite{blaizot06b,blaizot10,skokov10,herbst11,herbst12,herbst14,tripolt14}, which makes the approach convenient and reliable. The chiral invariant expansion technique, developed in \cite{fejos14}, has the numerical advantage that instead of calculating the effective potential in a three-dimensional grid, one may derive flow equations for one-dimensional coefficient functions. Furthermore, it should be noted that the FRG technique has not yet been applied widely to theories with nonvanishing $U_A(1)$ anomaly for $N_f=3$. A first analysis was presented in \cite{pawlowski98} for gauge theories, followed by \cite{jiang12} for the linear sigma model, also with the inclusion of quark degrees of freedom \cite{mitter14}. These studies provided important results on the effect of the anomaly factor; nevertheless, the flow of the anomaly coefficient and its temperature dependence in scalar theories have not been considered in the literature before. In this paper we are developing a method that besides providing the finite temperature flow of the `t Hooft term also realizes a resummation of a wide class of $U_A(1)$ breaking operators, as will be explained later. We emphasize that building phenomenology upon the scheme presented here is beyond the scope of the paper. In this study we are searching for an answer for the relevance of the mesonic fluctuations in regard to the induced temperature dependence of the anomaly factor. Explicit symmetry breaking terms representing finite quark masses are not introduced; nevertheless, we clarify the role of the six-point invariant which has not been investigated before.

The paper is organized as follows. In Sec. II, we introduce the model and review the chiral invariant expansion technique. We go beyond the approximation presented in \cite{fejos14}, and clarify the role of the 6-point invariant of the theory. In Sec. III, we introduce the anomaly and derive the flow equation for the field dependent anomaly coefficient, which is at first analyzed analytically, and then, in Sec. IV the reader finds the details of its numerical solution. Sec. V is dedicated to conclusions.

\section{Flow equations and the chiral invariant expansion}

The model to be investigated is a field theory of a $3\times 3$ matrix field $\Phi$, defined as
\bea
\Phi=(\sigma^a+i\pi^a)\frac{\hat{\lambda}^a}{2}, \qquad (a=0...8),
\eea
where $\hat{\lambda}^a$ are the Gell-Mann matrices, $\hat{\lambda}^0=\sqrt{\frac23} {\bf 1}$, with the $\sigma^a$, $\pi^a$ coefficients being scalar and pseudoscalar fields, respectively. $\Phi$ serves as an order parameter of the chiral transition, and its fluctuations give account of the scalar and pseudoscalar meson nonets. Before we discuss the details of the $U_A(1)$ factor, we go through the anomaly free model and the approximation scheme to be employed. The Lagrangian is
\bea
\label{Eq:Lag}
{\cal L}&=&\partial^{\mu} \Phi^\dagger \partial_\mu \Phi - m^2 \Tr ( \Phi^\dagger\Phi) \nonumber\\
&-&\lambda_1 [\Tr (\Phi^\dagger \Phi)]^2- \lambda_2\Tr (\Phi^\dagger \Phi\Phi^\dagger \Phi),
\label{Eq:lag}
\eea
which is clearly invariant under chiral $U(3)\times U(3)$ transformations. We choose $m^2<0$ and $\lambda_2>0$ (and also $\lambda_1+3\lambda_2>0$), being necessary conditions that lead to the expected symmetry breaking pattern $U(3)\times U(3) \longrightarrow U(3)$, realized by $\Phi=v_0\hat{\lambda}^0/2$; see details in \cite{fejos13}. 

For obtaining the effective potential of the model, we employ the functional renormalization group method \cite{wetterich93,pawlowski07}. The central object in the formalism is the scale dependent effective action $\Gamma_k$, which includes fluctuations with momenta $q\gtrsim k$, obeying the flow equation
\bea
\partial_k \Gamma_k = \frac12 \Tr \int \left[(\Gamma_k^{(2)}+R_k)^{-1}\partial_k R_k\right],
\label{Eq:flow}
\eea
where $\Gamma_k^{(2)}$ is the second functional derivative of $\Gamma_k$ with respect to $\Phi$, and $R_k$ is an appropriately chosen regulator function. It can be easily shown that if the UV cutoff is denoted by $\Lambda$, the scale dependent $\Gamma_k$ functional interpolates between the classical action [$k=\Lambda$] and the quantum effective action [$k=0$]. 

In this study we use the local potential approximation, which is the leading order contribution of the derivative expansion; in other words the effective action at all scales is approximated as
\bea
\Gamma_k[\Phi] = \int d^4x \left(\partial_\mu \Phi(x)^\dagger \partial^\mu \Phi(x) - V_k(x;\Phi)\right),
\eea
where $V_k$ is called the scale dependent effective potential. We use Litim's regulator \cite{litim01}:
\bea
\label{Eq:reg}
R_k(p_0,{\bf p})=(k^2-{\bf{p}}^2)\Theta(k^2-{\bf{p}}^2),
\eea
which, at finite temperature $T$, leads to 
\bea
\partial_k V_k[\Phi]=\frac{k^4}{6\pi^2} T\sum_{\omega_j}\sum_i \frac{1}{\omega_j^2+k^2+m_i^2(k)}.
\label{Eq:flow_Vk}
\eea
There is a finite temperature sum over bosonic Matsubara frequencies $\omega_j=2\pi j T$, and another one corresponding to the excitation spectrum. Here $m^2_i(k)$ denotes the eigenvalues of the scalar and pseudoscalar mass matrices $m^2_{\sigma,ij}(k)=\partial^2V_k/\partial \sigma^i\partial \sigma^j$ and $m^2_{\pi,ij}(k)=\partial^2V_k/\partial \pi^i \partial \pi^j$, respectively.

Now we review the chiral invariant expansion technique developed in \cite{fejos14}. The most important observation is that the $V_k$ local potential must reflect the $U(3)\times U(3)$ symmetry of the theory, which means that its variables are actually group invariants:
\bea
V_k(\Phi)\equiv V_k(I_1,I_2,I_3),
\eea
where
\bea
I_1&=&\Tr ({\Phi}^\dagger {\Phi}), \quad I_2=\Tr [{\Phi}^\dagger {\Phi}-\Tr({\Phi}^\dagger {\Phi})/3]^2, \nonumber\\
&& \quad I_3=\Tr [{\Phi}^\dagger {\Phi}-\Tr({\Phi}^\dagger {\Phi})/3]^3.
\eea
Since vector symmetries cannot be broken spontaneously \cite{vafa84}, we expect $\Phi \sim {\bf 1}$ and hence only $I_1$ takes nonzero value in the vacuum. Motivated by this observation, one attempts to expand $V_k(I_1,I_2,I_3)$ around $V_k(I_1,0,0)\equiv U(I_1)$ to get
\bea
V_k(I_1,I_2,I_3) \approx U_k(I_1) + C_k(I_1) I_2 + D_k(I_1) I_3 + ...
\label{Eq:Vk_inv}
\eea
where $C_k(I_1)=\partial V_k/\partial I_2|_{I_2=I_3=0}$ and $D_k(I_1)=\partial V_k/\partial I_3|_{I_2=I_3=0}$. At the UV scale $\Lambda$, in accordance with (\ref{Eq:lag}), we identify the coefficient functions with the following combination of the bare coupling constants:
\bea
U_{\Lambda}=m^2_{\Lambda}I_1 + (\lambda_{1\Lambda}+\frac{\lambda_{2\Lambda}}{3})I_1^2, \quad C_\Lambda(I_1)=\lambda_{2\Lambda}, 
\label{Eq:Vk_par}
\eea
with $D_{\Lambda}\equiv 0$. Higher order contributions in expansion (\ref{Eq:Vk_inv}) are going to be neglected, but note that, this already goes beyond the previous attempt \cite{fejos14}. Our task is to derive flow equations for $U_k(I_1)$, $C_k(I_1)$ and $D_k(I_1)$. Note that, since the mass eigenvalues [$m^2_i(k)$] are not chiral invariants, it is not straightforward to make (\ref{Eq:flow}) compatible with (\ref{Eq:flow_Vk}). A detailed description can be found in \cite{fejos14}; here we just sketch the procedure shortly.

\begin{figure*}
\begin{center}
\raisebox{0.05cm}{
\includegraphics[keepaspectratio,width=0.345\textwidth,angle=270]{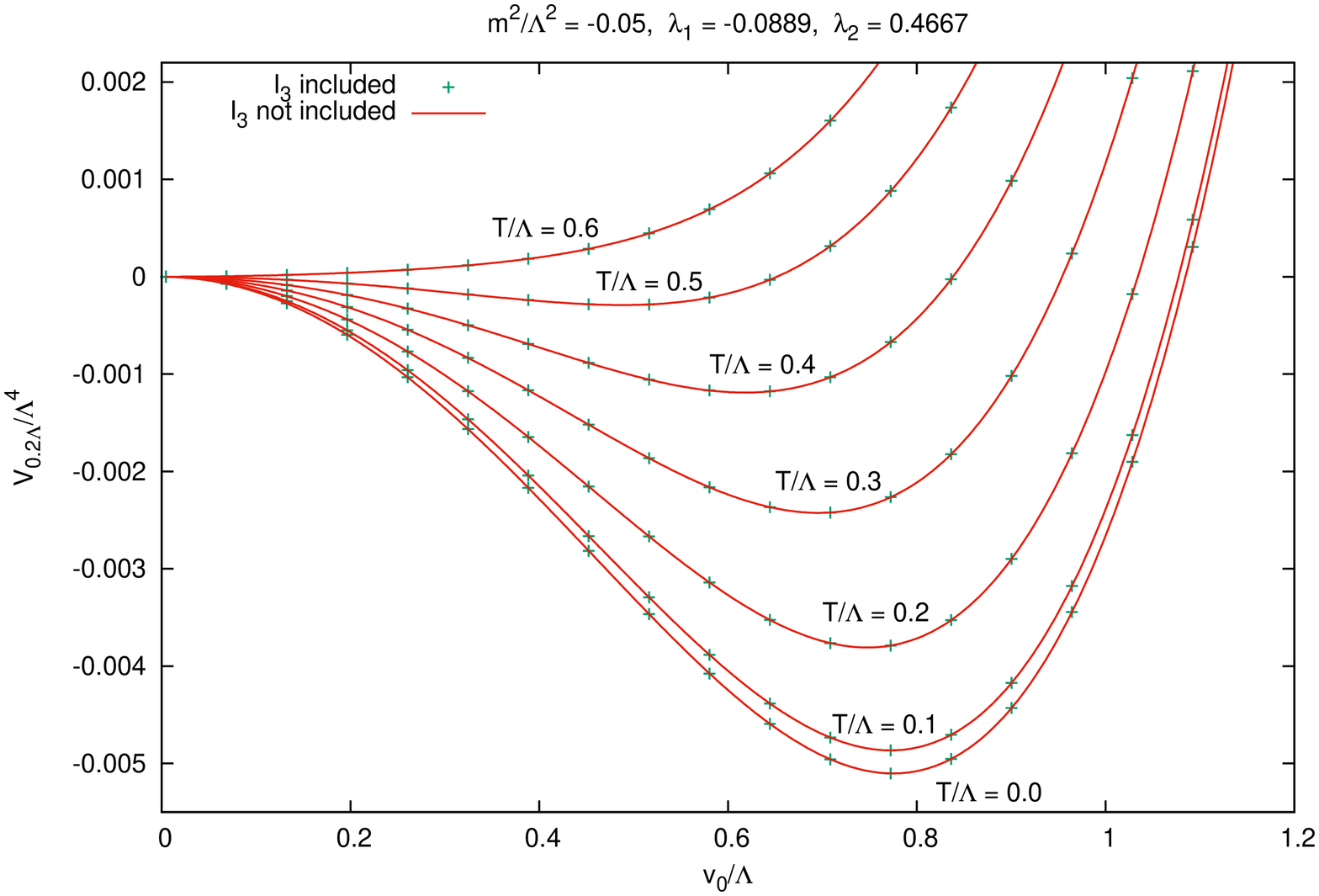}
\includegraphics[keepaspectratio,width=0.345\textwidth,angle=270]{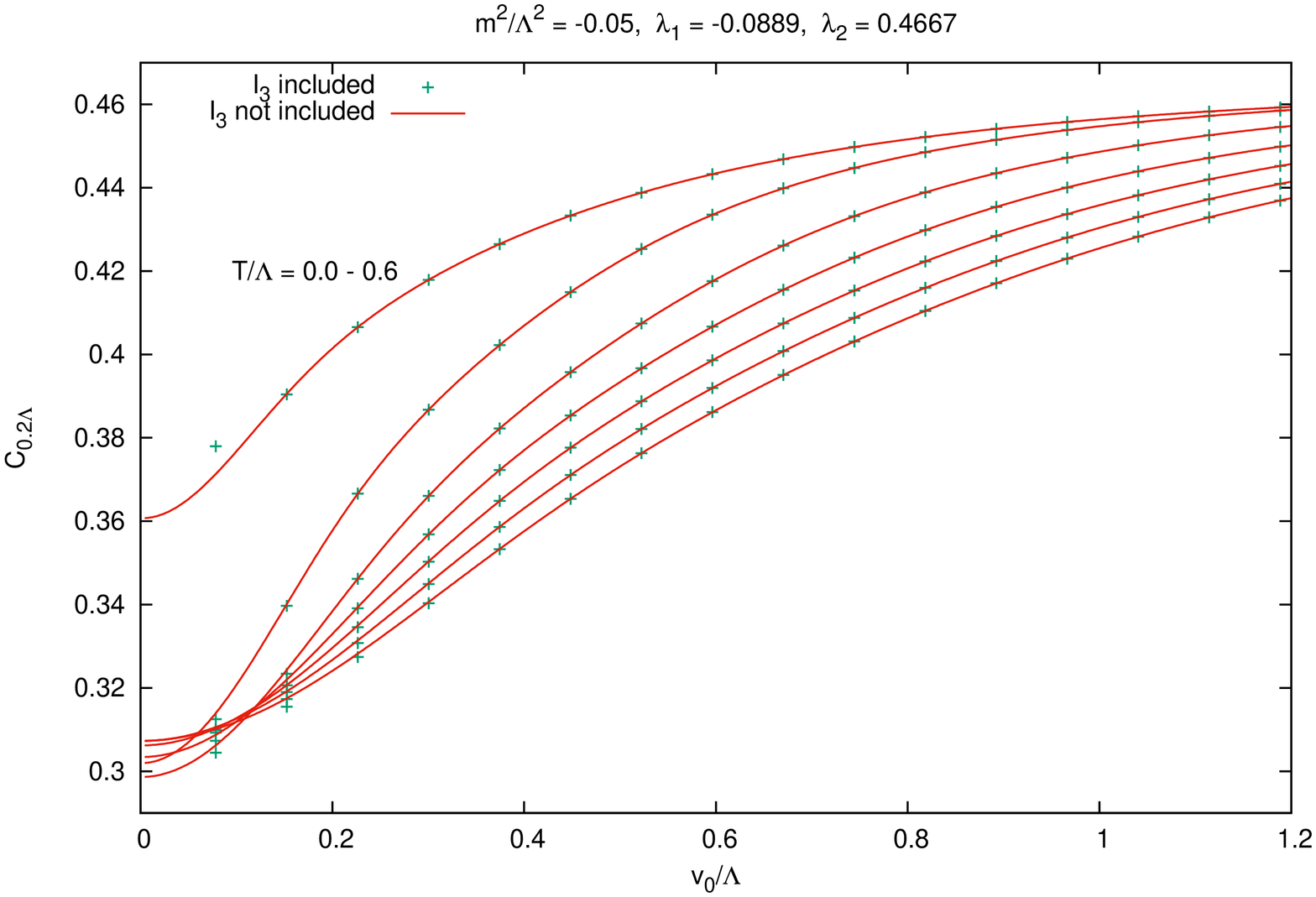}}
\vspace{0.5cm}\caption{Demonstration of the effect of the third invariant ($I_3$), as a function of the temperature. The plots show that $I_3$ has no effect on the effective potential, and only a moderate effect on the next-to-leading (NLO) order coefficient $C_k$ for small $v_0$ values. The temperature is varied between $T/\Lambda=0-0.6$ and increases on the left (decreases on the right) from bottom  to top.}
\label{fig1}
\end{center}
\end{figure*}

As already stressed, in the case of the symmetry breaking realized by $\Phi=v_0\hat{\lambda}^0/2 \sim {\bf 1}$, $I_2$ and $I_3$ vanishes; therefore, if we are to derive flow equations for the respective coefficient functions [i.e., $C_k(I_1)$ and $D_k(I_1)$], we have to assume the existence of a more general condensate. We choose to include an infinitesimal piece proportional to the matrix $\hat{\lambda}^8$: 
\bea
\Phi=v_0\hat{\lambda}^0/2+v_8\hat{\lambda}^8/2.
\label{Eq:v0v8}
\eea
Since
\begin{subequations}
\bea
I_1|_{v_0,v_8}&=&\frac{v_0^2+v_8^2}{2}, \\
I_2|_{v_0,v_8}&=&\frac{v_8^2}{24}(v_8-2\sqrt{2}v_0)^2, \\
I_3|_{v_0,v_8}&=&\frac{v_8^3}{288}(v_8-2\sqrt{2}v_0)^3,
\eea
\end{subequations}
\begin{figure*}
\begin{center}
\raisebox{0.05cm}{
\includegraphics[keepaspectratio,width=0.345\textwidth,angle=270]{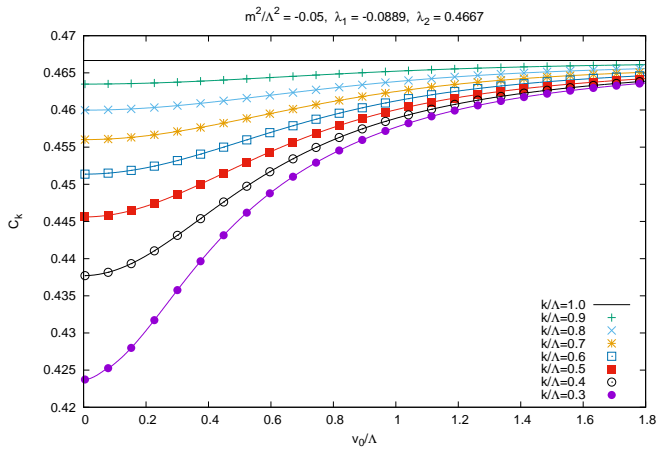}}
\includegraphics[keepaspectratio,width=0.345\textwidth,angle=270]{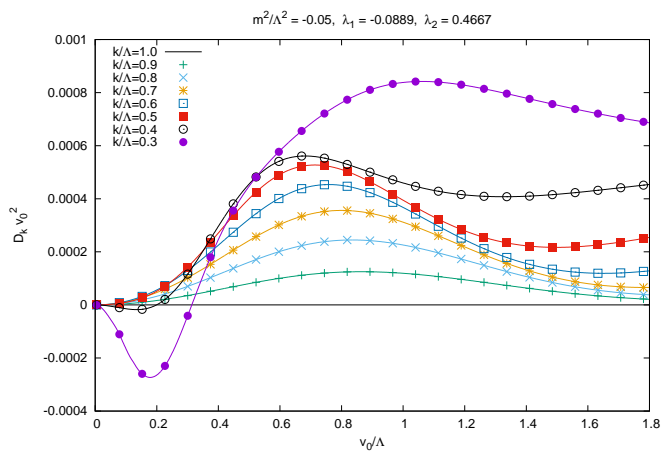}
\vspace{0.5cm} \caption{Comparison of next-to-leading (NLO) and next-to-next-to-leading (NNLO) coefficients of the chiral invariant expansion at zero temperature. Numerical values show that $D_k\cdot I_1 \ll C_k$, which means that the effect of the six-point invariant is negligible compared to that of the four-point invariant.}
\label{fig2}
\end{center}
\end{figure*}
the flow of each coefficient function $U_k(I_1)$, $C_k(I_1)$, and $D_k(I_1)$ can be easily identified; we just have to calculate the mass spectrum in the background (\ref{Eq:v0v8}) (see the appendix), substitute it into the right-hand side of (\ref{Eq:flow_Vk}), and expand it around $v_8=0$. Based on (\ref{Eq:Vk_inv}), the ${\cal O}(1)$ term will give $\partial_k U_k$, ${\cal O}(v_8^2)$ leads to the identification of $I_2$ and therefore $\partial_k C_k$, and finally ${\cal O}(v_8^4)$ provide $I_3$ and $\partial_k D_k$. These read as follows:
\begin{subequations}
\bea
\label{Eq:flow_Uk}
\partial_k&&U_k(I_1)=\frac{k^4T}{6\pi^2}\sum_{\omega_j}\Bigg[\frac{9}{\omega_j^2+E_\pi^2}+\frac{8}{\omega_j^2+E_{a_0}^2}+\frac{1}{\omega_j^2+E_\sigma^2}\Bigg], \nonumber\\
\\
\label{Eq:flow_Ck}
\partial_k&&C_k(I_1)=\frac{k^4T}{6\pi^2}\sum_{\omega_j}\Bigg[\frac{4(3C_k+2I_1C_k')^2/3}{(\omega_j^2+E_{a_0}^2)^2(\omega_j^2+E_\sigma^2)}\nonumber\\
&&+\frac{128C_k^5I_1^3/3}{(\omega_j^2+E_\pi^2)^3(\omega_j^2+E_{a_0}^2)^3}+\frac{24C_k\left(C_k-I_1C_k'\right)}{(\omega_j^2+E_{a_0}^2)^3}\nonumber\\
&&+\frac{4\left(3C_kC_k'I_1+4I_1^2C_k'+C_k(3C_k-2C_k''I_1^2)\right)/3}{(\omega_j^2+E_{a_0}^2)(\omega_j^2+E_\sigma^2)^2}\nonumber\\
&&+\frac{64C_k^3I_1^2(C_k-I_1C_k')/3}{(\omega_j^2+E_\pi^2)^2(\omega_j^2+E_{a_0}^2)^3}-\frac{48C_k^2I_1^2C_k'}{(\omega_j^2+E_\pi^2)(\omega_j^2+E_{a_0}^2)^3} \nonumber\\
&&+\frac{6C_k-17I_1C_k'}{(\omega_j^2+E_{a_0}^2)^2}\frac{1}{I_1}-\frac{6C_k+9I_1C_k'+2I_1^2C_k''}{(\omega_j^2+E_\sigma^2)^2}\frac{1}{I_1}\nonumber\\
&&+\frac{4C_k(6C_k+9I_1C_k'+2I_1^2C_k'')/3}{(\omega_j^2+E_{a_0}^2)(\omega_j^2+E_\sigma^2)^2}\Bigg],
\eea
\end{subequations}
where $\sigma, a_0$ and $\pi$ denote the scalar and pseudoscalar excitation spectrum, belonging to the breaking $U(3)\times U(3) \longrightarrow U(3)$. The energies are
\begin{subequations}
\label{Eq:energies}
\bea
E^2_\pi&=&k^2+M_\pi^2 \equiv k^2 + U'_k(I_1), \\
E^2_{a_0}&=&k^2+M_{a_0}^2\equiv k^2+U'_k(I_1)+\frac43 I_1C_k(I_1), \\
E^2_\sigma&=&k^2+M_\sigma^2\equiv k^2+U'_k(I_1)+2I_1U''_k(I_1),
\eea
\end{subequations}
with multiplicities $9\oplus 8\oplus 1$, respectively. The equation of $\partial_k D_k$ is too lengthy and we do not list it here explicity; it can be found in the appendix. All Matsubara sums appearing in the coupled flow equations can be performed analytically with the corresponding formulas also presented in the Appendix.

An important issue of the invariant expansion (\ref{Eq:Vk_inv}) is the investigation of its stability, which was not carried out in \cite{fejos14}. Going beyond this earlier work, i.e. the investigation of the flow equations of $U_k$ and $C_k$, now we are in a position to clarify the relevance of $D_k$. A few lines on the numerics can be found in Sec. V; here we just shortly review the results on the stability of the chiral invariant expansion.

In Fig. \ref{fig1}, we plot a typical solution of the temperature dependence of the effective potential at $k=0.2\Lambda$, with and without the six-point invariant term included; in other words we investigate how the solutions change if we drop the equation for $D_k$. One observes that the points are on top of each other, showing that the six-point invariant do not play any important role. This can be understood from Fig. \ref{fig2}, where we compare the relevance of $D_k$ with $C_k$. Based on dimensional grounds, if one is interested in the effects of the NLO ($C_k$) and NNLO ($D_k$) coefficients on the effective potential, one actually needs to compare $D_k\cdot I_1$ with $C_k$. The figure shows that since at the UV scale, $D_{k=\Lambda}$ is initiated as zero, its flow does not lead to a significant increase. Note that, this behavior is nontrivial, since there is no IR stable fixed point in the theory that would lead to scaling and (trivially) the suppression of nonrenormalizable couplings in the IR.

We expect that since the six-point invariant does not play a crucial role, none of the higher order Taylor coefficients have any effect on the solution either. With these findings the chiral invariant expansion has proven to be stable, and one can safely truncate the series at next-to-leading order.

\section{Inclusion of the $U_A(1)$ factor}

Implementation of the $U_A(1)$ anomaly is done via `t Hooft's determinant term. We add the following term into Lagrangian:
\bea
{\cal L}_{U_A(1)}=a \cdot (\det \Phi^\dagger + \det \Phi),
\eea
which explicitly breaks the $U_A(1)$ subgroup of the $U(3)\times U(3)$ chiral symmetry. The anomaly changes the spectrum to $8\oplus 1\oplus 7\oplus 1 \oplus 1$ [see also (\ref{Eq:anmass}) in the appendix], and the flow equation (\ref{Eq:flow_Vk}) generates all $U_A(1)$ breaking operators, such as, e.g., $\Tr(\Phi^\dagger \Phi)\left(\det \Phi^\dagger + \det \Phi\right)$ into the effective potential.

The effect of the anomaly on the effective potential can be formulated as
\bea
V_k=V_k&&|_{a=0}(I_1,I_2,I_3)\nonumber\\
\!\!\!&&+\sum_{i} A_k^{(i)}(I_1,I_2,I_3) (\det \Phi^\dagger + \det \Phi)^i,
\eea
where $V_k|_{a=0}$ is the anomaly free effective potential. Since the anomaly can only be carried by the $I_{\det}\equiv \det \Phi^\dagger + \det \Phi$ operator, the $A_k^{(i)}$ coefficients are $U(3)\times U(3)$ invariants, and thus one can apply the chiral invariant expansion on them as well:
\bea
A_k^{(i)}(I_1,I_2,I_3)=A_k^{(i)}(I_1,0,0)+\sum_{\{\alpha\}} \prod_{j=2,3} A^{(i)}_{k,\alpha}(I_1) I_j^{\alpha_j}.\nonumber\\
\eea
In this paper we only present a leading order analysis:
\bea
\label{Eq:Ak1}
V_k \approx V_k|_{a=0}+A_k^{(1)}(I_1,I_2,I_3) (\det \Phi^\dagger + \det \Phi),
\eea
and furthermore, we neglect the $I_2$ and $I_3$ dependence of $A_k$,
\bea
\label{Eq:Ak2}
A_k^{(1)}(I_1,I_2,I_3)\approx A_k^{(1)}(I_1,0,0)\equiv A_k(I_1).
\eea
Note that a field dependent $U_A(1)$ coefficient already goes beyond a simple perturbative renormalization group analysis of the anomaly. Taylor series of $A_k(I_1)$ around $I_1=0$ shows that an infinite resummation of operators $[\Tr(\Phi^\dagger \Phi)]^n (\det \Phi^\dagger + \det \Phi)$ realizes in the system.

As already announced in the introduction, in the literature the anomaly coefficient ($a$) is somewhat controversially linked to the topological susceptibility. Here we treat it as a temperature independent parameter and investigate how thermal fluctuations of mesons affect the strength of $A_k(I_1)$.

\begin{figure*}
\begin{center}
\raisebox{0.05cm}{
\includegraphics[keepaspectratio,width=0.345\textwidth,angle=270]{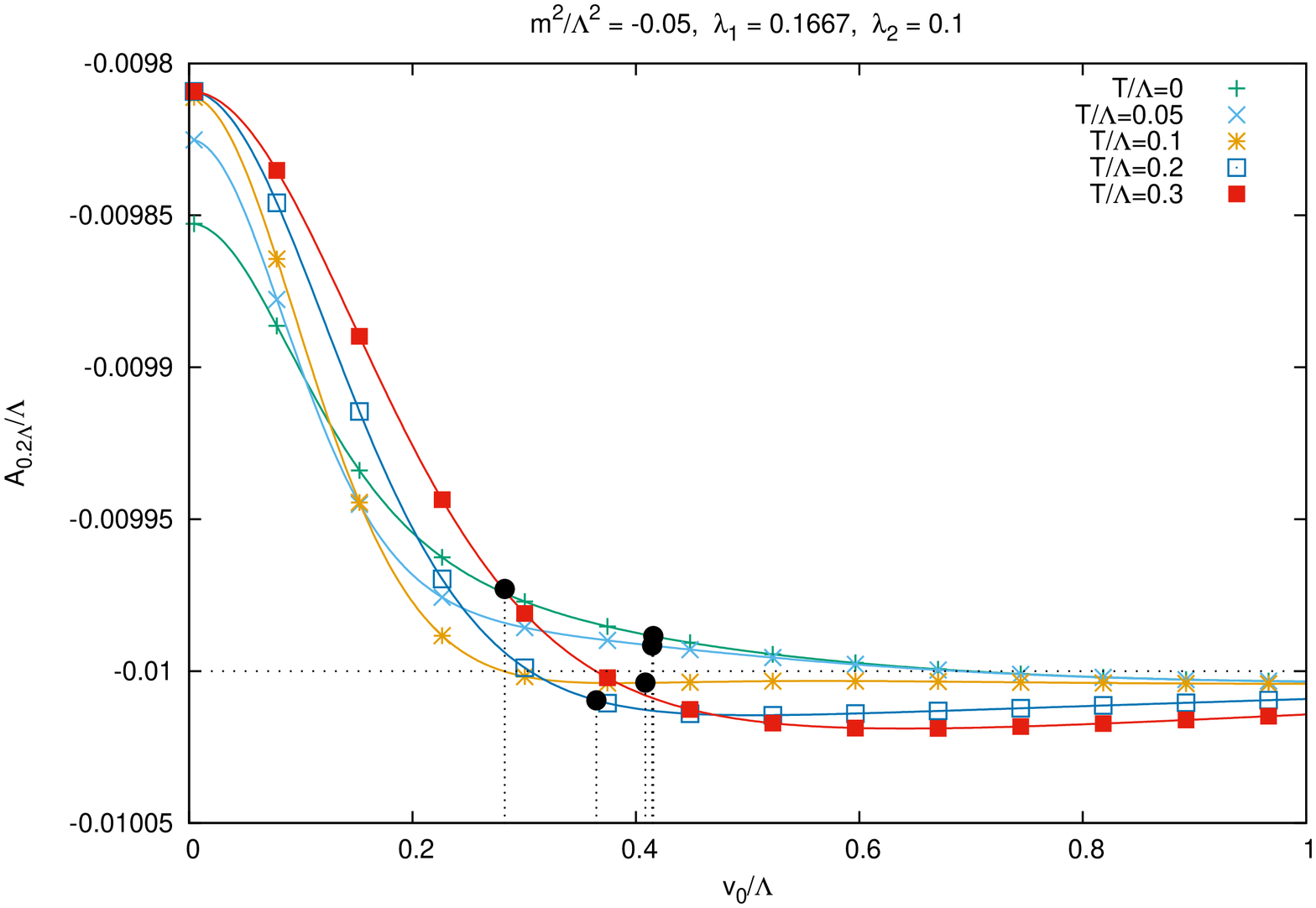}
\includegraphics[keepaspectratio,width=0.345\textwidth,angle=270]{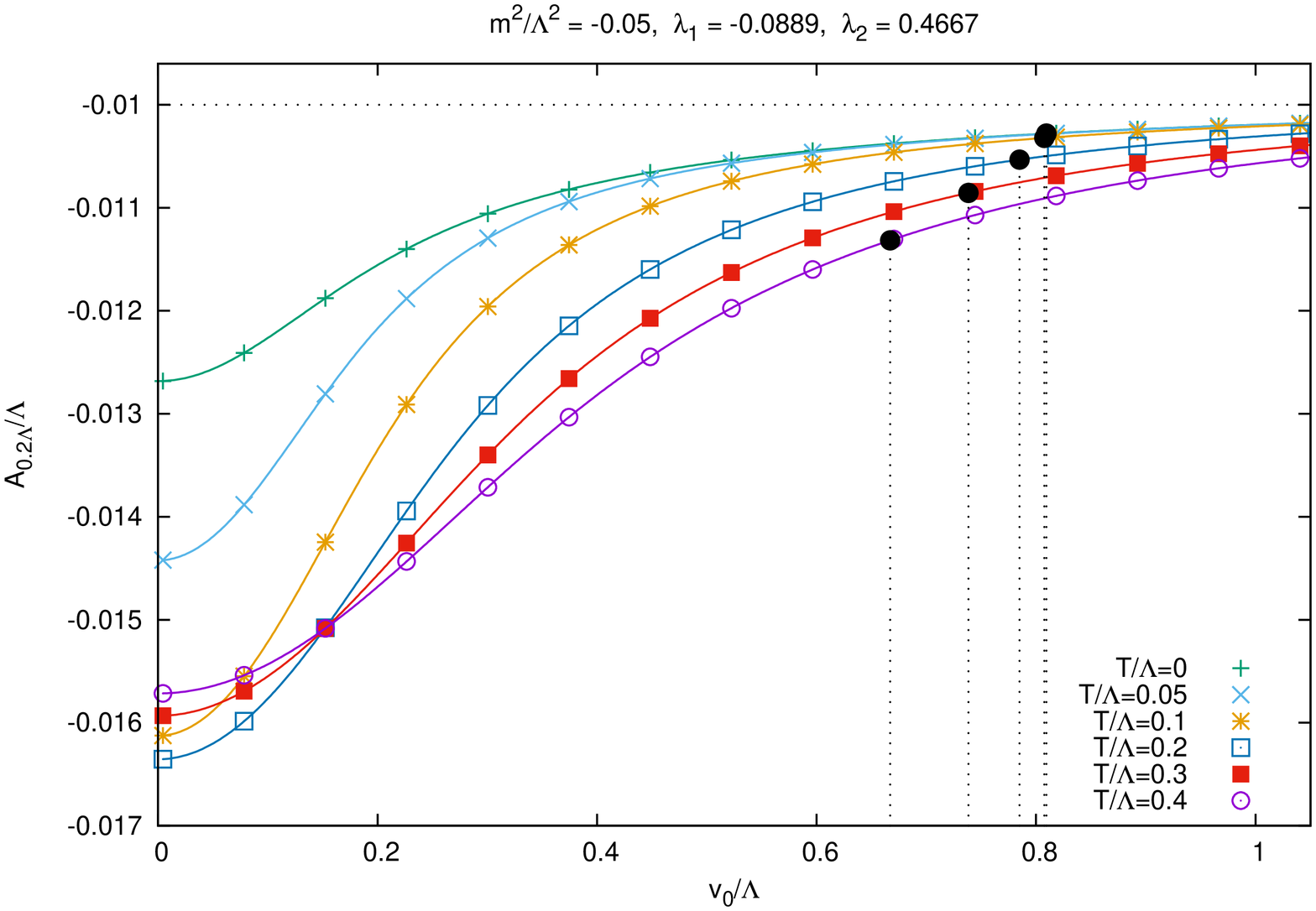}}
\vspace{0.5cm}\caption{Temperature dependence of the anomaly coefficient at $k=0.2\Lambda$. The plots show that the rough analytical estimate is qualitatively correct, depending on the sign of $(\lambda_{2\Lambda}-\lambda_{1\Lambda}$) the anomaly can either weaken or strengthen due to mesonic fluctuations. Horizontal dotted lines represent the anomaly coefficient at the UV scale, and the circle-shaped spots show the field values that minimize the effective potential, together with the corresponding anomaly strengths.}
\label{fig3}
\end{center}
\end{figure*}

In order to obtain an approximate flow equation compatible with (\ref{Eq:Ak1})-(\ref{Eq:Ak2}), we have to expand (\ref{Eq:flow_Vk}) around the zero anomaly configuration, i.e., $A_k(I_1)\equiv 0$. The reader is referred to the appendix for formulas of the mass matrices and the necessary derivatives of invariants. The zeroth order of the expansion gives the already obtained flow equations for $U_k$, $C_k$, and $D_k$, while the next-to-leading order terms combine into a term proportional to $I_{\det}$. The coefficient of $I_{\det}$ becomes the scale derivative of $A_k(I_1)$. We arrive at
\bea
\label{Eq:flow_Ak}
\partial_k A_k(I_1)&=&\frac{k^4}{6\pi^2}T\sum_{\omega_j} \Bigg[-\frac{9A_k'}{(\omega_j^2+E_\pi^2)^2}-\frac{9A_k}{I_1(\omega_j^2+E_\pi^2)^2}\nonumber\\
&-&\frac{8A_k'}{(\omega_j^2+E_{a_0}^2)^2}+\frac{12A_k}{I_1(\omega_j^2+E_{a_0}^2)^2}\nonumber\\
&-&\frac{3A_k}{(\omega_j^2+E_\sigma^2)^2I_1}+\frac{7A_k'}{(\omega_j^2+E_\sigma^2)^2}+\frac{2I_1A_k''}{(\omega_j^2+E_\sigma^2)^2}\Bigg]. \nonumber\\
\eea
This equation can now be solved numerically given that the functions $U_k$, $C_k$ and $D_k$ are known.

Before we present the full numerical results, let us solve (\ref{Eq:flow_Ak}) using the assumption of $V_k$ taking the form of the classical (bare) potential, with $k$-dependent, field independent couplings. For constant $A_k$, at high enough temperatures, around $I_1 \approx 0$ (\ref{Eq:flow_Ak}) simplifies:
\bea
\partial_k \log A_k = -\frac{2k^4 T}{3\pi^2} \frac{8C_k-3U_k''}{(k^2+U_k')^3}.
\eea
The effective potential in our approximation, $V_k\equiv U_k(I_1)+C_k(I_1)\cdot I_2+D_k(I_1)\cdot I_3+A_k(I_1)\cdot I_{\det}$ can be parametrized as [see also (\ref{Eq:Vk_par})]
\label{Eq:Ans}
\bea
U_k(I_1)&=&m_k^2I_1+(\lambda_{1k}+\frac{\lambda_{2k}}{3})I_1^2, \nonumber\\
C_k(I_1)&=&\lambda_{2k}, \qquad D_k(I_1)\equiv 0, \nonumber\\
A_k(I_1)&=&a_k,
\eea
from which at high temperature
\bea
\label{Eq:ancond}
a_k=a_{\Lambda}e^{\frac{4T}{\pi^2}\int_k^\Lambda d\kappa \frac{\kappa^4}{(\kappa^2+U_\kappa')^3}(\lambda_{2\kappa}-\lambda_{1\kappa})}.
\eea
This result shows that within the field independent coupling approximation, and with the assumption that the flow is monotonic, the condition of decreasing anomaly strength as a function of the temperature can be formulated already at bare level:
\bea
\label{Eq:anom_rel}
\lambda_{2\Lambda}<\lambda_{1\Lambda}.
\eea

\section{Numerical results}

Equation (\ref{Eq:ancond}) might not survive exactly in the numerics, but it turns out to be a qualitatively correct approximation. The applied method for solving the coupled flow equations (\ref{Eq:flow_Uk})-(\ref{Eq:flow_Ck}), (\ref{Eq:flow_Dk}), and (\ref{Eq:flow_Ak}) has been the same as in \cite{fejos14}. At all scales ($k$) the $U_k(I_1)$, $C_k(I_1)$, $D_k(I_1)$, and $A_k(I_1)$ functions are stored on a grid, typically in the region $[0,2]$, with a step size of $\sim 10^{-3}$, and all dimensionful units are measured in terms of the UV cutoff $\Lambda$. The flow equations are integrated with the Runge-Kutta method with a typical step size of $\Delta k \sim 10^{-5}$. We also emphasize that, due to numerical stability the flows have to be stopped \cite{fejos14} around the scale where the potential is gradually becoming convex \cite{oraif86,berges02,litim97}. As shown in \cite{fejos14}, critical temperatures and the corresponding discontinuities of the order parameter (in case of first order transitions) can be obtained by extrapolation to $k=0$ from the $k>0$ results. Note that, in the present study however, we decided to stop the flows uniformly at $k=0.2\Lambda$, since we are only interested in the tendency of the evolution of the temperature and field dependent anomaly, which can perfectly be seen around $k\approx 0.2\Lambda$. Extrapolated results to $k=0$ with physical parametrization of the model will be reported elsewhere.

In Fig. \ref{fig3}, we show the $U_A(1)$ coefficient function at two distinct points of the parameter space that lead to either anomaly strengthening or weaking. We observe that already at zero temperature, quantum fluctuations develop a structure for $A_k(I_1)$ showing that a field independent approximation is not appropriate. It can also be seen that the full numerical solution is in accordance with (\ref{Eq:anom_rel}); at $I_1 \approx 0$, for large $\lambda_{2}$ the anomaly is getting weaker as a function of the temperature, while if $\lambda_{1}$ dominates, it is strengthening. It is interesting to see that for larger $I_1$ values the former statement is not necessarily true anymore: even though at small values of the field the anomaly weakens, if $I_1$ is larger, it might also strengthen at the same time. In case of large $\lambda_2$, we observe that there might be a maximum value for the $U_A(1)$ factor as a function of the temperature, followed by a moderate decrease.

The vacuum state is of course always determined by the effective potential; therefore, in Fig. \ref{fig3}. we also indicate the field values that minimize the complete effective potential (i.e., $V_k$). This shows that a field independent anomaly coefficient [belonging to the leading order of the Taylor expansion of $A_k(I_1)$ around $I_1\approx 0$] is indeed a crude approximation; at some temperatures the actual strength of the anomaly in the minimum may differ up to even $\sim$50$\%$ from the value of it in the origin. Based on this observation, several possibilities can arise as the temperature is raised. The anomaly can either weaken or strengthen (more or less) monotonically, or it can increase and discontinuously drop at the transition point. The latter can be smoothened by explicit symmetry breaking terms representing finite quark masses, or by the initial strength of the anomaly itself. The mapping of this issue will be reported elsewhere.

We note that recent tree-level parametrizations of the model \cite{lenaghan00}, or extensions of the model \cite{parganlija13,kovacs15}, show that $\lambda_2>\lambda_1$. It is important to stress that one should not draw any final conclusions on the behavior of the anomaly based on these findings, since the value of the bare coupling constants strongly depend on the cutoff scale and regularization, and also on the employed approximations. There are other recent parametrizations of the model, using the FRG formalism, where $\lambda_1 \sim \lambda_2$ \cite{kamikado15}.

Finally let us point out that since the $I_1$ invariant specifies an $O(18)$ combination of the fields [$I_1 \sim (\sigma^a)^2+(\pi^a)^2$], roughly speaking $\lambda_1$ characterizes the orthogonal nature of the theory, while $\lambda_2$ shows its deviation from it. Based on this argument, one can reformulate (\ref{Eq:anom_rel}) as a requirement of the theory to be closer to the $O(18)$ model, if one is to seek decreasing anomaly factor as a function of the temperature.
\vspace{0.2cm}
\section{Conclusions}

In this study we have analyzed the three flavor linear sigma model with the local potential approximation of the functional renormalization group in $3+1$ dimensions. The chiral invariant expansion introduced in \cite{fejos14} has been developed to the case of nonzero $U_A(1)$ anomaly, and its stability has also been checked explicitly. At first, by deriving the flow equation of the next-to-next-to-leading coefficient of the expansion we have found that the chiral invariant expansion does appear to be stable; there is no need to go beyond next-to-leading order. Furthermore, we derived the flow equation of the field dependent $U_A(1)$ anomaly factor, and found qualitatively that the first quartic coupling has to dominate the second for decreasing anomaly as a function of the temperature. 

It has turned out that mesonic fluctuations are capable of describing anomaly weakening, even without an implicit temperature dependence of the `t Hooft determinant coupling. Adding explicit symmetry breaking terms, and parametrizing the model with low energy mesonic spectrum would allow us to investigate temperature dependence of the anomaly and spectrum itself in the physical point. This represents a future study to be reported in the near future.

\section*{Acknowledgements}
The author thanks T. Hatsuda and K. Kamikado for useful comments and discussions. This work was supported by the Foreign Postdoctoral Research Program of RIKEN.
\newline

\makeatletter
\@addtoreset{equation}{section}
\makeatother 

\renewcommand{\theequation}{A\arabic{equation}} 

\begin{widetext}

\appendix 
\section{Mass matrices and group invariants}  

In this appendix we give all the necessary formulas to calculate the scalar and pseudoscalar mass matrices $m^2_{\sigma,ij}(k)=\frac{\partial^2 V_k}{\partial \sigma^i \partial \sigma^j}$ and $m^2_{\pi,ij}(k)=\frac{\partial^2 V_k}{\partial \pi^i \partial \pi^j}$. They are needed to obtain the flow equations for $U_k$, $C_k$ and $D_k$, and ultimately for $A_k$.
\begin{subequations}
\bea
m_{\sigma,ij}^2&=&\frac{\partial I_1}{\partial \sigma^i}\frac{\partial I_1}{\partial \sigma^j}\Big(U_k''(I_1)+C_k''(I_1)\cdot I_2+D_k''(I_1)\cdot I_3\Big)
+\frac{\partial^2 I_1}{\partial \sigma^i\partial \sigma^j}\Big(U_k'(I_1)+C_k'(I_1)\cdot I_2+D_k'(I_1)\cdot I_3\Big)\nonumber\\
&+&\Big(\frac{\partial I_1}{\partial \sigma^i}\frac{\partial I_2}{\partial \sigma^j}+\frac{\partial I_1}{\partial \sigma^j}\frac{\partial I_2}{\partial \sigma^i}\Big)C_k'(I_1)
+\Big(\frac{\partial I_1}{\partial \sigma^i}\frac{\partial I_3}{\partial \sigma^j}+\frac{\partial I_1}{\partial \sigma^j}\frac{\partial I_3}{\partial \sigma^i}\Big)D_k'(I_1),
+\frac{\partial^2 I_2}{\partial \sigma^i\partial \sigma^j}C_k(I_1)+\frac{\partial^2 I_3}{\partial \sigma^i\partial \sigma^j}D_k(I_1), \\
m_{\pi,ij}^2&=&\frac{\partial I_1}{\partial \pi^i}\frac{\partial I_1}{\partial \pi^j}\Big(U_k''(I_1)+C_k''(I_1)\cdot I_2+D_k''(I_1)\cdot I_3\Big)
+\frac{\partial^2 I_1}{\partial \pi^i\partial \pi^j}\Big(U_k'(I_1)+C_k'(I_1)\cdot I_2+D_k'(I_1)\cdot I_3\Big)\nonumber\\
&+&\Big(\frac{\partial I_1}{\partial \pi^i}\frac{\partial I_2}{\partial \pi^j}+\frac{\partial I_1}{\partial \pi^j}\frac{\partial I_2}{\partial \pi^i}\Big)C_k'(I_1)
+\Big(\frac{\partial I_1}{\partial \pi^i}\frac{\partial I_3}{\partial \pi^j}+\frac{\partial I_1}{\partial \pi^j}\frac{\partial I_3}{\partial \pi^i}\Big)D_k'(I_1)
+\frac{\partial^2 I_2}{\partial \pi^i\partial \pi^j}C_k(I_1)+\frac{\partial^2 I_3}{\partial \pi^i\partial \pi^j}D_k(I_1).
\eea
\end{subequations}
If we include the `t Hooft determinant, and the corresponding $A_k(I_1)\cdot (\det \Phi +\det \Phi^\dagger)\equiv A_k(I_1) \cdot I_{\det}$ term in the scale dependent effective potential, the mass matrices get corrected by
\begin{subequations}
\label{Eq:anmass}
\bea
\Delta m_{\sigma,ij}^2&=&\frac{\partial I_1}{\partial \sigma^i}\frac{\partial I_1}{\partial \sigma^j}A_k''(I_1)\cdot I_{\det}+\frac{\partial^2 I_1}{\partial \sigma^i\partial \sigma^j}A_k'(I_1)\cdot I_{\det} 
+\Big(\frac{\partial I_1}{\partial \sigma^i}\frac{\partial I_{\det}}{\partial \sigma^j}+\frac{\partial I_1}{\partial \sigma^j}\frac{\partial I_{\det}}{\partial \sigma^i}\Big) A_k'(I_1)
+\frac{\partial^2 I_{\det}}{\partial \sigma^i \partial \sigma^j}A_k(I_1), \\
\Delta m_{\pi,ij}^2&=&\frac{\partial I_1}{\partial \pi^i}\frac{\partial I_1}{\partial \pi^j}A_k''(I_1)\cdot I_{\det}+ \frac{\partial^2 I_1}{\partial \pi^i\partial \pi^j}A_k'(I_1)\cdot I_{\det} 
+\Big(\frac{\partial I_1}{\partial \pi^i}\frac{\partial I_{\det}}{\partial \pi^j}+\frac{\partial I_1}{\partial \pi^j}\frac{\partial I_{\det}}{\partial \pi^i}\Big) A_k'(I_1)
+\frac{\partial^2 I_{\det}}{\partial \pi^i \partial \pi^j}A_k(I_1).
\eea
\end{subequations}

Invariants $I_1$ and $I_2$, and their first derivatives in the two-component background are the following:
\begin{subequations}
\bea
I_1|_{v_0,v_8}&=&\frac{v_0^2+v_8^2}{2}, \hspace{7.45cm} I_2|_{v_0,v_8}=\frac{v_8^2}{24}(v_8-2\sqrt{2}v_0)^2, \\
\frac{\partial I_1}{\partial \sigma^a}\bigg|_{v_0,v_8}&=&v_0\delta^{a0}+v_8\delta^{a8}, \hspace{6.25cm} \frac{\partial I_1}{\partial \pi^a}\bigg|_{v_0,v_8}=0, \\
\frac{\partial I_2}{\partial \sigma^a}\bigg|_{v_0,v_8}&=&\left(\frac{2v_0v_8^2}{3}-\frac{1}{3\sqrt{2}}v_8^3\right)\delta^{a0}+\left(\frac{2v_0^2v_8}{3}-\frac{v_0v_8^2}{\sqrt{2}}+\frac{v_8^3}{6}\right)\delta^{a8}, \hspace{0.41cm} \frac{\partial I_2}{\partial \pi^a}\bigg|_{v_0,v_8}=0.
\eea
\end{subequations}
The second derivatives are
\begin{subequations}
\bea
\frac{\partial^2 I_1}{\partial \sigma^a \partial \sigma^b}\bigg|_{v_0,v_8}&=&\delta^{ab}, \hspace{7.3cm} \frac{\partial^2 I_1}{\partial \pi^a \partial \pi^b}\bigg|_{v_0,v_8}=\delta^{ab}, \\
\frac{\partial^2 I_2}{\partial \sigma^a \sigma^b}\bigg|_{v_0,v_8}&=&
\begin{cases}
\frac{2}{3}v_8^2,  \hspace{4.5cm} \ife \hspace{0.1cm} a=b=0\\
-\frac{v_8^2}{\sqrt{2}}+\frac{4}{3}v_0v_8,  \hspace{3.0cm} \ife \hspace{0.1cm} a=0,\hspace{0.1cm} b=8 \hspace{0.1cm} \orr \hspace{0.1cm} a=8,\hspace{0.1cm} b=0\\
\frac{2}{3}v_0^2+\frac{v_8^2}{2}-\sqrt{2}v_0v_8,  \hspace{2.1cm} \ife \hspace{0.1cm} a=b=8\\
\frac{2}{3}v_0^2+\frac{v_8^2}{6}+\sqrt{2}v_0v_8, \hspace{2.1cm} \ife \hspace{0.1cm} a=b=1,2,3\\
\frac{2}{3}v_0^2+\frac{v_8^2}{6}-\frac{1}{\sqrt{2}}v_0v_8, \hspace{2.15cm} \ife \hspace{0.1cm} a=b=4,5,6,7\\
0, \hspace{4.9cm}  \els\\
\end{cases}\\
\frac{\partial^2 I_2}{\partial \pi^a \pi^b}\bigg|_{v_0,v_8}&=&
\begin{cases}
0, \hspace{4.9cm} \ife \hspace{0.1cm} a=b=0\\
-\frac{v_8^2}{3\sqrt{2}}+\frac{2}{3}v_0v_8, \hspace{2.9cm} \ife \hspace{0.1cm} a=0,\hspace{0.1cm} b=8 \hspace{0.1cm} \orr \hspace{0.1cm} a=8,\hspace{0.1cm} b=0\\
\frac{v_8^2}{6}-\frac{\sqrt{2}}{3}v_0v_8, \hspace{3.18cm} \ife \hspace{0.1cm} a=b=8\\
-\frac{v_8^2}{6}+\frac{\sqrt{2}}{3}v_0v_8, \hspace{2.88cm} \ife \hspace{0.1cm} a=b=1,2,3\\
\frac{5}{6}v_8^2-\frac{1}{3\sqrt{2}}v_0v_8, \hspace{2.83cm} \ife \hspace{0.1cm} a=b=4,5,6,7\\
0, \hspace{4.9cm}  \els.\\
\end{cases}
\eea
\end{subequations}
Furthermore, invariants $I_3$ and $I_{\det}$, and their first derivatives are
\begin{subequations}
\bea
I_3|_{v_0,v_8}&=&\frac{v_8^2}{288}(v_8-2\sqrt{2}v_0)^3, \hspace{6.4cm} I_{\det}|_{v_0,v_8}=\frac{v_0^3}{3\sqrt{6}}-\frac{v_0v_8^2}{2\sqrt{6}}-\frac{v_8^3}{6\sqrt{3}}, \\
\frac{\partial I_3}{\partial \sigma^a}\bigg|_{v_0,v_8}&=&-\frac{v_8^3(v_8-2\sqrt2 v_0)^2}{24\sqrt2}\delta^{a0}-\frac{v_8^2}{48}(v_8-2\sqrt2 v_0)^2(\sqrt2 v_0-v_8)\delta^{a8}, \hspace{0.4cm} \frac{\partial I_3}{\partial \pi^a}\bigg|_{v_0,v_8}=0, \\
\frac{\partial I_{\det}}{\partial \sigma^a}\bigg|_{v_0,v_8}&=&\frac{2v_0^2-v_8^2}{2\sqrt{6}}\delta^{a0}-\frac{v_8(\sqrt{2}v_0+v_8)}{2\sqrt{3}}\delta^{a8}, \hspace{4.05cm} \frac{\partial I_{\det}}{\partial \pi^a}\bigg|_{v_0,v_8}=0.
\eea
\end{subequations}
Finally, the second derivatives are
\begin{subequations}
\bea
\frac{\partial^2 I_3}{\partial \sigma^a \sigma^b}\bigg|_{v_0,v_8}&=&
\begin{cases}
\frac{v_8^3}{\sqrt6}(v_8-2\sqrt2 v_0),  \hspace{5.5cm} \ife \hspace{0.1cm} a=b=0\\
-\frac{v_8^2}{48}(24\sqrt2 v_0^2-32v_0v_8+5\sqrt2 v_8^2),  \hspace{3.05cm} \ife \hspace{0.1cm} a=0,\hspace{0.1cm} b=8 \hspace{0.1cm} \orr \hspace{0.1cm} a=8,\hspace{0.1cm} b=0\\
\frac{v_8}{48}(-16\sqrt2 v_0^3+48v_0^2v_8-20\sqrt2 v_0v_8+5v_8^3),  \hspace{1.6cm} \ife \hspace{0.1cm} a=b=8\\
\frac{v_8}{48}(16\sqrt2 v_0^3+16v_0^2v_8+4\sqrt2 v_0v_8^2-5v_8^3), \hspace{2.05cm} \ife \hspace{0.1cm} a=b=1,2,3\\
\frac{v_8}{48}(-8\sqrt2 v_0^3+16v_0^2 v_8 -5\sqrt2 v_0v_8^2+v_8^3), \hspace{2.15cm} \ife \hspace{0.1cm} a=b=4,5,6,7\\
0, \hspace{7.8cm}  \els\\
\end{cases}\\
\frac{\partial^2 I_3}{\partial \pi^a \pi^b}\bigg|_{v_0,v_8}&=&
\begin{cases}
0,  \hspace{7.8cm} \ife \hspace{0.1cm} a=b=0\\
-\frac{v_8^2}{24\sqrt2}(8 v_0^2-4\sqrt2 v_0v_8+v_8^2),  \hspace{3.65cm} \ife \hspace{0.1cm} a=0,\hspace{0.1cm} b=8 \hspace{0.1cm} \orr \hspace{0.1cm} a=8,\hspace{0.1cm} b=0\\
\frac{v_8^2}{48}(8 v_0^2-4\sqrt2v_0v_8+v_8^2),  \hspace{4.3cm} \ife \hspace{0.1cm} a=b=8\\
-\frac{v_8^2}{48}(8v_0^2-4\sqrt2 v_0 v_8+v_8^2), \hspace{4.0cm} \ife \hspace{0.1cm} a=b=1,2,3\\
\frac{v_8^2}{48}(4v_0^2-11\sqrt2 v_0 v_8 +5v_8^2), \hspace{3.95cm} \ife \hspace{0.1cm} a=b=4,5,6,7\\
0, \hspace{7.75cm}  \els\\
\end{cases}\\
\frac{\partial^2 I_{\det}}{\partial \sigma^a \sigma^b}\bigg|_{v_0,v_8}&=&
\begin{cases}
\sqrt{\frac23}v_0,  \hspace{7.0cm} \ife \hspace{0.1cm} a=b=0\\
-\frac{v_8}{\sqrt6},  \hspace{7.2cm} \ife \hspace{0.1cm} a=0,\hspace{0.1cm} b=8 \hspace{0.1cm} \orr \hspace{0.1cm} a=8,\hspace{0.1cm} b=0\\
-\frac{v_0}{\sqrt6}-\frac{v_8}{\sqrt3},  \hspace{6.3cm} \ife \hspace{0.1cm} a=b=8\\
-\frac{v_0}{\sqrt6}+\frac{v_8}{\sqrt3}, \hspace{6.3cm} \ife \hspace{0.1cm} a=b=1,2,3\\
-\frac{v_0}{\sqrt3}-\frac{v_8}{2\sqrt3}, \hspace{6.15cm} \ife \hspace{0.1cm} a=b=4,5,6,7\\
0, \hspace{7.7cm}  \els\\
\end{cases}\\
\frac{\partial^2 I_{\det}}{\partial \pi^a \pi^b}\bigg|_{v_0,v_8}&=&
\begin{cases}
-\sqrt{\frac23}v_0,  \hspace{6.75cm} \ife \hspace{0.1cm} a=b=0\\
\frac{v_8}{\sqrt6},  \hspace{7.45cm} \ife \hspace{0.1cm} a=0,\hspace{0.1cm} b=8 \hspace{0.1cm} \orr \hspace{0.1cm} a=8,\hspace{0.1cm} b=0\\
\frac{v_0}{\sqrt6}+\frac{v_8}{\sqrt3},  \hspace{6.55cm} \ife \hspace{0.1cm} a=b=8\\
\frac{v_0}{\sqrt6}-\frac{v_8}{\sqrt3}, \hspace{6.55cm} \ife \hspace{0.1cm} a=b=1,2,3\\
\frac{v_0}{\sqrt6}+\frac{v_8}{2\sqrt3}, \hspace{6.4cm} \ife \hspace{0.1cm} a=b=4,5,6,7\\
0. \hspace{7.75cm}  \els
\end{cases}
\eea
\end{subequations}

\renewcommand{\theequation}{B\arabic{equation}} 

\section{NNLO flow equation}

Here we present the flow of the coefficient function $D_k(I_1)$; for explanation see (\ref{Eq:flow_Vk}), (\ref{Eq:Vk_inv}), and (\ref{Eq:energies}).
\bea
\label{Eq:flow_Dk}
\partial_k D_k=\frac{k^4}{6\pi^2}&&\!\!\!\!\!\!\!T\sum_{\omega_j} \Bigg[\frac{c_1}{(\omega_j^2+E_\sigma^2)^2}+\frac{c_2}{(\omega_j^2+E_\sigma^2)^2(\omega_j^2+E_{a_0}^2)}+\frac{c_3}{(\omega_j^2+E_{a_0}^2)^2}+\frac{c_4}{(\omega_j^2+E_{a_0}^2)^2(\omega_j^2+E_{\sigma}^2)}\nonumber\\
&+&\frac{c_5}{(\omega_j^2+E_{a_0}^2)^3}+\frac{c_6}{(\omega_j^2+E_{a_0}^2)^4}+\frac{c_7}{(\omega_j^2+E_{a_0}^2)^3(\omega_j^2+E_{\pi}^2)}+\frac{c_8}{(\omega_j^2+E_{\pi}^2)^2}\nonumber\\
&+&\frac{c_9}{(\omega_j^2+M_{a_0}^2)^3(\omega_j^2+M_\pi^2)^2}+\frac{c_{10}}{(\omega_j^2+E_\pi^2)^3}+\frac{c_{11}}{(\omega_j^2+E_{a_0}^2)^3(\omega_j^2+E_{\pi}^2)^3}+\frac{c_{12}}{(\omega_j^2+E_\pi^2)^4}\Bigg],
\eea
where the $c_j$ coefficients are the following:
\begin{subequations}
\bea
c_1=&&(12C_k^2D_kI_1-24C_kD_kI_1^2C_k'-48D_kI_1^3C_k'^2+20C_k^2I_1^2D_k'+48I_1^3C_kC_k'D_k'-8C_k^2I_1^3D_k''-243C_k^2U_k''\nonumber\\
&&-90C_kD_kI_1U_k''-324I_1C_kC_k'U_k''-180I_1^2D_kC_k'U_k''-108I_1^2C_k'^2U_k''+48I_1^2C_kD_k'U_k''  -72I_1^3C_k'D_k'U_k''\nonumber\\
&&+24I_1^3C_kD_k''U_k''-135I_kD_kU_k''^2-117I_1^2D_k'U_k''^2-18I_1^3D_k'U_k'')/I_1^2(2C_k-3U_k'')^2, \\
c_2=&&-18C_k^2/I_1-24C_kC_k'-8I_1C_k'^2, \\
c_3=&&(60C_k^2D_kI_1+216C_k^2I_1C_k'+96C_kD_kI_1^2C_k'+192D_kI_1^3C_k'^2-416C_k^2I_1^2D_k'-192C_kI_1^3C_k'D_k'+972C_k^2U_k''\nonumber\\
&&+36C_kD_kI_1U_k''+648I_1C_kC_k'U_k''+720D_kI_1^2C_k'U_k''+432I_1^2C_k'^2U_k''+816C_kI_1^2D_k'U_k''+288I_1^3C_k'D_k'U_k''\nonumber\\
&&+783D_kI_1U_k''^2+486I_1C_k'U_k''^2-288I_1^2D_k'U_k''^2)/4I_1^2(2C_k-3U_k'')^2, \\
c_4=&&-18C_k^2/I_1-24C_kC_k'-8I_1C_k'^2 \\
c_5=&&(18C_k^3-4C_kD_k^2I_1^2+180I_1C_k^2C_k'+96I_1^2D_kC_kC_k'+48I_1^2C_kC_k'^2+32D_kC_k'I_1^3+135C_k^2U_k'' \nonumber\\
&&+108C_kD_kI_1U_k'' +6D_k^2I_1^2U_k''-54I_1C_kC_k'U_k'')/I_1(C_k-3U_k''/2), \\
c_6=&&2(27C_k^3+54C_k^2D_kI_1+36C_kD_k^2I_1^2+8D_k^3I_1^3), \\
c_7=&&72I_1C_k^2C_k', \\
c_8=&&-27D_k/4I_1-27C_k'/2I_1-9D_k', \\
c_9=&&-32C_k^4I_1+32I_1^2C_k^3C_k', \\
c_{10}=&&18C_k(D_k+C_k/I_1), \\
c_{11}=&&-64C_k^5I_1^2/3, \\
c_{12}=&&-6C_k^3.
\eea
\end{subequations}

\renewcommand{\theequation}{C\arabic{equation}} 

\section{Matsubara sums}
The Matsubara sums appearing in the flow equations have the following form:
\bea
S(i,j)=\sum_{\omega_m} \frac{1}{(\omega_m^2+E_1^2)^i (\omega_m^2+E_2^2)^j}.
\eea
First, we calculate $S(1,0)$ and $S(1,1)$:
\bea
S(1,0)&=&\frac{\coth(E_1/2T)}{2E_1},\\
S(1,1)&=&\frac{1}{2E_1E_2}\frac{E_1\coth(E_2/2T)-E_2\coth(E_1/2T)}{E_1^2-E_2^2}.
\eea
The rest can be obtained by differentiation:
\bea
S(n>1,0)&=&\frac{(-i)^{n-1}}{(n-1)!}\frac{\partial^{n-1}S(1,0)}{\partial(E_1^2)^{n-1}},\\
S(n>1,m>1)&=&\frac{(-i)^{n-1}}{(n-1)!}\frac{(-i)^{m-1}}{(m-1)!}\frac{\partial^{n-1}}{\partial(E_1^2)^{n-1}}\frac{\partial^{m-1}S(1,1)}{\partial(E_2^2)^{m-1}}.
\eea
\end{widetext}

\end{document}